\documentclass[showpacs,preprintnumbers,amssymb,aps,prb]{revtex4}
\newcommand{\beq}{\begin{equation}}
\newcommand{\eeq}{\end{equation}}

\begin{document}
\title{Spontaneous symmetry breaking in a system of strongly interacting
multicomponent fermions \\(electrons with spin and conducting
nanotubes)}
\author{V. V. Afonin, V. L. Gurevich}
\affiliation{A. F. Ioffe Institute of the Russian Academy of Sciences, 194021
Saint Petersburg, Russia}
\author{V. Yu. Petrov}
\affiliation{Theory Division, Saint Petersburg, Nuclear Physics
Institute of the Russian Academy of Sciences}
\begin{abstract}
We calculate the ground state wave functions for a systems of
multicomponent interacting fermions. We show that it describes the
state with spontaneously broken chiral symmetry. In the limit of
an  infinitely strong interaction it turns into a  phase with a
finite density of chiral complexes. The number of particles
constituting a complex depends on the number of fermion
components. For example, in the case of two component electrons
(spin) the condensate is built of four-particle complexes
consisting of two ,,right`` electrons and two ,,left`` holes with
the opposite spins.
\end{abstract}
\maketitle
\section{Introduction and  discussion of results}
\label{I} Advances in semiconductor technology have renewed
interest in the properties of one-dimensional (1D) electron
systems. It is well known that the electron-electron (e-e)
interaction alters properties of 1D system qualitatively. To gain
a better understanding of the problem one should try to clarify
the physical question of a primary importance: what is the nature
of the ground state of this system? In order to get the answer one
usually studies the ground state for 1D interacting fermions using
the ,,density-density`` correlation functions. However this
information is not direct. As it has been shown the correlation
functions of the problem contain the terms that decay very slowly.
Usually they are interpreted in the following way: the
oscillations with the Fermi momentum ($p_F$) doubled are related
to the Peierls instability (Refs.~\cite{V},\cite{E}) while the
oscillations behaving as $4p_F$ are interpreted as a marginal
Wigner crystal~\cite{Shul}.

In Ref.~\cite{AP} the wave function of the ground state
of spinless fermions was constructed for the exactly soluble
Tomonaga -- Luttinger model. It has been shown
that at sufficiently low temperatures the system should be in
the state that has nothing in common with a system undergoing the
Peierls transition. It is a state with a spontaneously broken
chiral symmetry. There is a long-range order in the electron
system. In the limit of infinitely strong interaction, at low
temperatures a condensate of finite density is formed. It
consists of neutral (exciton-like) pairs of a right electron and
a left hole or {\em vice versa}. The uniqueness of a one
dimensional (1D) system impels one to consider a phase transition
of the 2nd kind in a channel of a finite length $L_{\Vert}$. The
point is that the temperature of the phase transition vanishes as
$1/L_{\Vert}$ as it should be. Note that actually the phase transition
temperature need not be too small. For the length $L_{\Vert} \sim
10^{-4}$~cm it should be about $1$ K. In other words,
consideration of the limit $L_{\Vert} \to \infty $ cannot be sufficient
for description of the modern experiment.

Traditionally many-component fermions in the 1D systems have
been extensively discussed in the literature. The interest was
going to rouse by separation the spatial and spin degrees of
freedom~\cite{V},\cite{E}. In the present paper we are going
to discuss the form of the ground state wave function for this
case. As the variables are separable one can expect that the
ground state wave function is a direct product of two factors
where one of the factors, describing the spinless component, coincides
with the ground state wave function of spinless fermions. In the present
paper we show that in fact one has an entirely different
situation. Namely, the most correlated state is the state where
the total spin vanishes. However, for $n$-component electrons
this state consists not of pairs (as in the case of the spinless
fermions) but of pointlike neutral complexes containing $2n$ particles
and having a chirality $\pm2n$.

For the ordinary electron system, $n=2$, one has the complexes
consisting of two right electrons and two left holes with opposite
spins.  For the conducting nanotubes $n=4$
(Refs.\onlinecite{OD},\onlinecite{NT},\onlinecite{K}), and the
complexes consists of eight particles. The complexes with a
smaller number of particles can have a nonzero spin but their
correlation is much weaker.  For example, for $n=2$ the spin phase
can be realized only as a Kosterlitz-Thouless one, and for the
limit of infinitely strong interaction the density tends to zero
as $1/\sqrt{L_{\Vert}}$. By contrast, the spinless phase has  for
this limit a finite density.

This situation is typical for many field-theoretical models with
Adler anomaly (this is the case for Luttinger model as well). It
is known that in such models the  new fermion interaction
("t'Hooft interaction" \cite{thooft}) can appear with a vortex
which is a product over all components of fermions. In many cases
t'Hooft interaction leads to the spontaneous breakdown of chiral
invariance with 2n fermion order parameter. In particular, this is
a case for multicomponent Schwinger model\cite{smilga} and, as we
shall see, for the Luttinger model in the limit of infinitely
strong interaction.

For last model the most correlated state is built out of complexes
each of them having the maximal possible number of degrees of
freedom. This state differs qualitatively from the result of a
common treatment, i. e.  marginal Wigner crystal~(\cite{Shul}).
Instead of the phase transition of almost  first kind one gets the
phase transition of almost  second kind. In order to manifest
breakdown of the chiral symmetry in the Luttinger liquid we have
calculated exactly the wave function of the ground state in this
model, Eq.~(\ref{s20}), and demonstrated explicitly that its
symmetry is less than the original symmetry of the Hamiltonian.
(This is the definition of spontaneous symmetry breaking). These
are basically the main results of the paper.

For one component fermion system pointlike complexes with more
than 2 particles are forbidden by Pauli principle. This is not the
case for the multicomponent fermion. For this reason, expression
for the ground state wave function Eq.~(\ref{s20}) is much more
cumbersome and the calculations are more involved.  Therefore we
will present the results only for a short range potential in the
limit of infinitely strong interaction.  At the same time, in our
case the physical picture is quite similar to that of the one
component fermions. For instance, had one taken into account
corrections in the reciprocal strength of interaction one would
have gotten a Kosterlitz-Thouless phase for spinless complexes
too. The strength of the interaction going up, it would be
gradually transformed into a state of a definite condensate
density.

The paper is organized as follows. In Section~\ref{I} we have
given a brief review of our results. Section~\ref{G} contains
discussion of the main difference between the multicomponent
problem and spinless one in regard of the theoretical description
and most essential steps of calculation. In Section~\ref{CN} we
give arguments concerning the possibility to apply the theory to
nanotubes. Appendix is devoted to a derivation of some
intermediate results.

\section{Description of approach and derivation of the main results}
\label{G}

Our starting point will be the usual Tomonaga --- Luttinger
Hamiltonian $H$ (see, for instance, Refs.\onlinecite{M},
\onlinecite{E}) for a system of interacting electrons without
regard of the backscattering.  For the  case where the interaction
does not change the electron spin the Hamiltonian $H$ can be
expressed through the density of the right ($R$) and left ($L$)
electrons
$$\varrho\left( x \right)_{\alpha }=\varrho_{R,\alpha }\left( x
\right)+ \varrho_{L,\alpha }\left( x  \right)$$ (the spin index
$\alpha$ equals $ \pm $ for the spin $\pm1/2$ respectively) as

 $$H = \sum_{\alpha }\int dx \left[ \hat
\Psi^{\dag}_{R,\alpha } \left( x \right) v_F\left( -i \partial_x
\right) \hat \Psi_{R,\alpha } \left( x  \right) + \hat
 \Psi^{\dag}_{L,\alpha } \left( x \right) v_F i\partial_x \hat
\Psi_{L,\alpha } \left( x  \right) \right]$$
\begin{equation}   +\int dx dy \varrho \left( x  \right) V\left(
x-y \right) \varrho\left( y  \right) .\nonumber \label{g1}
\end{equation}
Here $v_F$ is the Fermi velocity and
\begin{equation}
\hat \Psi _{\alpha }\left( x\right)=\exp\left( ip_F x\right)\hat
\Psi _{R,\alpha }\left( x \right) +\exp\left( -ip_F x\right)\hat
\Psi_{L,\alpha } \left( x \right) ,
\label{g2}
\end{equation}
while $\varrho$ is the total density. For simplicity, we assume
that the interaction is spin-independent as well (it is not too
essential for calculations). Now we introduce the electron and
hole operators in the usual way

\begin{center}
\begin{eqnarray}
\hat \Psi_{\left( R,L \right) \alpha } \left( x  \right)= \int
\limits_0^{\infty} \frac{dp}{2 \pi}
\left( \exp\left( \pm ipx \right)\hat a_{\left( R,L \right)
\alpha } \left( p  \right) +
\exp\left( \mp ipx \right)\hat b_{\left( R,L \right) \alpha
}^{\dag} \left( p  \right)\right)=\\
\hat a_{\left( R,L \right) \alpha }\left( x  \right) +
\hat b_{\left( R,L \right) \alpha }^{\dag}\left(x  \right) \nonumber
\label{g2'}
\end{eqnarray}
\end{center}
Here $\hat a_{ R, +}\left( x \right)$ is the operator of
annihilation of an electron with spin $+1/2$ while $\hat b_{R, + }
\left(x \right)$ is the operator of annihilation of a hole with
spin $-1/2$. To write such interaction via functional integral it
is necessary to introduce one Bose field $\Phi$ and apply a
version of the Hubbard --- Stratonovich identity~\cite{Hub}:
\begin{eqnarray}
\exp \left[\frac{i}{2 }\int_0^Tdt \int^\infty_{-\infty}
\frac{dp}{2\pi} V \left( p \right) \varrho \left( p,t \right)
\varrho \left( -p,t  \right) \right] =\nonumber\\
\frac{1}{{\cal N }}\int {\cal D}\Phi\exp \left[ \frac{i}{2}
 \int_0^T dt \int^\infty
_{-\infty}\frac{dp}{2\pi} \Phi\left( p,t \right) \Phi
\left( -p,t \right) V^{-1} \left( p  \right) \right. \nonumber\\
- \left. \frac{i}{2}\int_0^T dt \int^\infty _{-\infty}
\frac{dp}{2\pi} \left( \varrho \left( p,t \right) \Phi\left( -p,t
\right) + \varrho
 \left( -p,t  \right)
\Phi\left( p,t  \right)\right) \right] . \label{g3}
\end{eqnarray}
Here $V \left( p  \right)$ is the Fourier-transform of the e-e interaction
and the normalization factor is equal to
\begin{equation}
{\cal N } = \int {\cal D}\Phi \exp \left[ \frac{i}{2} \int_0^T dt
\int^\infty _{-\infty}\frac{dp}{2\pi} \Phi\left( p,t \right)
\Phi\left( -p,t \right) V^{-1} \left( p  \right) \right] .
\label{g4}
\end{equation}
This identity shows that the theory with e-e interaction is
equivalent to a theory of non-interacting electrons in an external
field $\Phi$. (Had one taken a theory with spin-dependent
interaction he has to introduce two independent Bose fields. The
calculation would be more cumbersome but the physical picture
would be the same.)

Calculation of the ground state wave function is given in detail
in Ref.~\cite{AP}. It is based on calculation of the evolution
operator for the electrons
\begin{equation} S\left( T \right) = \sum_{m,n} |n>
<n|\exp{(-iHT)}|m> <m| .
\label{s1}
\end{equation} Here $|n>$ are the
exact wave functions of the Hamiltonian $H$ in the secondary
quantization representation, $T$ is the observation time.  $S(T)$
determines the evolution of an arbitrary initial wave function ($<m|$)
from the time $t=0$ up to final states  $|n>$ (at $t=T$). (Henceforth
we imply that the Schr\"odinger representation for operator with
time-dependent wave-functions is used.)

Eq.~(\ref{s1}) suggests the general method to obtain the wave
functions. One has to calculate first the evolution operator and
present it as a sum of time-dependent exponents. The coefficients
in front of these exponents are products of the exact wave functions
and their complex conjugates. In order to extract the ground state
wave function one has to take the limit $T\to\infty$ (we add
infinitesimal imaginary part to the energy). Proceeding to
Euclidean time ($T\rightarrow -i/\Theta$) we see that evolution
operator determines the density matrix for the equilibrium system
at a non-zero temperature $\Theta$.

As for the case of spinless electrons (see~\cite{AP} for the
details), one can present the evolution operator for the
electrons in an external field as
\begin{equation}
 {\hat S \left( \Phi
\right)}=\exp {\left( {\cal S}_0 + \log\right[{\rm Det}\,\Phi
\left(T\right) \left] \right)}|F><F|.
\label{s2}
\end{equation}

Before integration over the fields $\Phi$ the equation for ${\hat
S \left( \Phi\right)}$ undergoes some changes in comparison with
the spinless case. They amount to appearance of a factor $n$, the
number of components of the electron wave function, in the
equation describing the quantum fluctuations in the electron
system under the action of the field $\, \Phi \left( T \right)$
(in the spin case $n=2$)
\begin{equation}
\ln\left[{\rm Det}\, \Phi\left(T\right)\right] = -\frac{n}{4\pi}
\int_0^T dt dt_1\int _{-\infty}^\infty\frac{dp}{2\pi} \Phi \left(
 -p,t  \right) \Phi \left( p,t_1  \right) |p| \exp\left[ -i|p|v_F
 |t-t_1|  \right] .
\label{g5}
\end{equation}
The operator structure of Eq.~(\ref{s2}) is determined by the
second part of action, ${\cal S}_0$. Here one should take into
account the spin index
\begin{eqnarray} {\cal S}_0 =
\sum_{i=R,L; \alpha} \int dx dx'  \left[  \hat b_{i,\alpha} \left( x '
\right) G_{i}^0\left( x',0;x,\epsilon \right) \hat a_{i,\alpha}\left( x
\right) + \hat a^{\dag }_{i,\alpha}\left( x ' \right) G_{i}^0\left(
x',T;x,T-\epsilon\right) \hat b^{\dag }_{i,\alpha} \left( x  \right)
\right.\nonumber\\ -\left. \hat a_{i,\alpha}^{\dag}\left( x ' \right)
G_{i}^0\left( x',T;x,0\right) \hat a_{i,\alpha}\left( x  \right) - \hat
b_{i,\alpha}\left( x ' \right) G_{i}^0\left( x',0;x,T\right) \hat
b_{i,\alpha}^{\dag }\left( x  \right) \right] , \label{g7}
\end{eqnarray}
where $G_i^0$ is the spin-independent free particle Green
function
\begin{equation}
G_{R,L}^0 \left( x,t;x_1,t_1\right) =\frac{1}{2\pi i}
\left[ v_f\left( t-t_1\right) \mp \left( x-x_1\right)
-i\delta\, {\rm sign}\left( t-t_1\right) \right] ^{-1}.
\label{s6}
\end{equation}

Behaviour of the multicomponent fermions in the external
field is quite similar to the one-component case.
Essential complication, however, appears after the integration of the
operator ${\hat S \left( \Phi \right)}$ over $\Phi$ with
the weight~(\ref{g3}).

Two points should be indicated.
\begin{itemize}
\item The coefficient $n$ in Eq.~(\ref{g5}) that enters the
equation for the action. Due to this coefficient, after
calculation of the integral over $\Phi$ the analytical structure
of the resulting expression becomes much more complicated. As a
result, in the integrals defining wave functions of the
multiparticle complexes, one gets cuts instead of simple poles, as
in one component case. This leads to rather cumbersome complex
wave functions. In particular, complexes are non-local.
\item The
nonlocal property brings about essential enhancement of the
various electron states. A number of electron states is forbidden
for one component electrons due to Pauli principle. In contrast to
this, in the multicomponent case the number of connected diagrams
becomes infinite. This makes the expression for the ground state
wave function rather cumbersome. However the very fact of
existence of chiral phase for the infinitely strong interaction
still persists.

\end{itemize}

Because of non-Gaussian form of the final functional integral
it is impossible to perform the integration in Eq.~(\ref{g5})
over  $\Phi(x,t)$ in a closed form but it is possible
to obtain an arbitrary term of the
evolution operator expanding it in ${\cal S}_0^n$. It will be
sufficient in order to get the ground state wave function as we will
have the integral of a Gaussian type that can be
easily calculated. After doing the final integration over $\Phi$
the final recipe of calculation of the evolution operator can be
written as a sum  of the following terms
\begin{equation}
{\cal S}_0^n \left(  \hat a_{\left( R,L \right)}
, \hat b_{\left( R,L \right)},....  \right)  \exp{\left( {\cal
S}^{\rm eff}_n \right)} |F><F|.
\label{g6}
\end{equation}

Here $|F>$ is the Fermi ''sphere``  while the terms in ${\cal
S}_0^n$ determine the operator structure of the wave functions, i.
e. all the possible particle configurations as a result of their
interaction. (\ref{g6}) is a sort of symbolic expression. Indeed,
the analytic equation for the effective action ${\cal S}^{\rm
eff}_n $ in the $n$th term of expansion depends explicitly on the
particle configuration in the preexponential factor $\left( {\cal
S}_0\right)^n $. Naturally, it is different for different terms.
Note that the evolution operator (\ref{s1}) is determined in such
a manner that the initial state expressed through the electron and
hole annihilation operators and the final state determined through
creation operators are given at different times. It means that
{\em  during  calculating the evolution operator} one should
consider the operators $\hat a_{R,L}^{\dag}\left( x \right)$ and
$\hat a_{R,L}\left( y  \right)$ as anticommuting.

In order to write the expression for ${\cal S}^{\rm eff}_n$, we
introduce the following notation for the coordinates
of the electron-hole creation-annihilation operators:
\begin{enumerate}
\item We will denote by $x$ ($y$) the coordinates of the
right (left) particles.
\item We will put a tilde on the coordinates related to annihilation
operators (the initial state): the coordinates of creation
operators (the final state) will have no tilde.
\item We will prime the hole coordinates.
\end{enumerate}

The effective action differs from the action for the one-component
fermions only by a factor and in the limit of strong interaction
$$\frac{V\left( p\right)}{\pi v_F} \gg 1 $$ is equal
$$ {\cal
S}_{\rm eff}=-\frac{\pi}{nL}\sum_{m\ne 0}\frac{1}{|p_m|} \left[   {\cal
R }_f \left( -p,x_1\dots\right){\cal R }_f \left( p,x_1\dots
\right) + {\cal R }_ i \left( -p,\tilde x_1\dots\right) {\cal R
}_i \left( p,\tilde x_1\dots \right)\right]$$ \begin{equation}
-\frac{2\pi}{n L}\sum_{m\ne 0}\frac{1}{|p_m|}\exp\left(\frac{-|p_m|v_f}
{\Theta}\right) {\cal R}_f \left( -p, x_1\dots \right) {\cal R }_i
\left( p,\tilde x_1\dots \right) , \label{s7} \end{equation}

The extra factor $n$ is the number of the fermion components.
It appears because in the Luttinger model the excitation
spectrum is~\cite{TS}:
\begin{equation}
\omega_p = |p|v_F \sqrt{ 1 +\frac{nV\left( p  \right)}{\pi v_F}}.
\label{s8}
\end{equation}
Eq.~(\ref{s7}) is valid in the temperature region
$$
\Theta_{\rm chiral}\ll \Theta\ll \Theta_c=
\omega_{ \left( p_{\rm min} \right)}.
$$

The right-hand side of the last inequality is the energy of
excitations with a minimal momentum (for periodic boundary
conditions $p_{\rm min}  = 2\pi / L_{\Vert} $) while  $\Theta_{\rm
chiral} = |p_{\rm min}|v_F$ is the degeneracy temperature
of the ground state. In this temperature region one can neglect
the energy difference between the states of different chiralities.
This means that the ground states with different chiralities
becomes degenerate.

The origin of this inequality has been discussed in detail
in Ref.~\cite{AP}. It is not sensitive to the number of
the wave function components. For the temperatures
$\Theta\ll \Theta_{\rm chiral}$ the last term in Eq.~(\ref{s7})
should be omitted. Then the corresponding equation will be valid
for the low temperatures too (see detailed discussion in
Ref.~ \cite{AP}).

The functions ${\cal R }_{i,f} \left( p,\tilde x_1\dots\right)$
in the equation~(\ref{s7}) depend explicitly on the electron ($x..$) and
hole ($\tilde x ...$) coordinates in the preexponential factor. These functions
are given by
\begin{eqnarray}
{\cal R}_f\left( p,
x\dots \right) = \sum_{x\dots;x'\dots;y\dots;y'\dots } \theta \left( p
\right)\left[ \exp \left( ipx \right)\right.\nonumber \\ -
\left.\exp \left( ipx' \right) \right] +  \theta \left( -p
\right)\left[ \exp \left( ipy \right) - \exp \left( ipy' \right)
\right] ,\nonumber\\ {\cal R}_i\left( p,\tilde x...  \right)=
\sum_{\tilde x\dots;\tilde x'\dots;\tilde y\dots;\tilde y'\dots} \theta
\left( -p \right)\left[ \exp \left( ip\tilde x \right) \right.\\
-\left. \exp \left( ip\tilde x' \right) \right] +  \theta \left(
p \right)\left[ \exp \left( ip\tilde y \right) - \exp \left(
ip\tilde y' \right) \right] \nonumber \label{s9}
\end{eqnarray}

To obtain the complete expression for the ground state wave
function one has to consider all the complexes and separate the
connected parts out of them. It is not necessary, however, since,
according to the general theorem, the complete wave function is
the exponent of the connected complexes~\cite{FS}.

For one-component fermions there is only one possible 2-particle
connected complex. This is not true for multi-component case, many
of the scattering channels are possible, so the number of
connected diagrams is infinite. In principle, one can calculate
the exact wave function of any given complex taking the Gaussian
integral in $\Phi$. Unfortunately, it is not enough to present the
whole wave function of the system in a closed form, but actually
in order to prove the existence of symmetry breaking  we do not
needed it. To be sure of the fact it is sufficient  to prove that
the wave function symmetry is less than Hamiltonian. In order to
do this one should analyze the simplest connected diagrams
bringing about a spontaneous breaking of the Hamiltonian symmetry.
The rest terms either have the symmetry of the Hamiltonian or
describe the scattering of the simplest correlated complexes and
also violate the chiral symmetry.

Now we embark on analysis of the simplest diagrams of
the evolution operator for the electrons having a spin.
We will begin from the temperature region
$\Theta_{\rm chiral}\gg \Theta$. In this case the action
and therefore the evolution operator factorize so that
one can consider explicitly the ground state wave function
$|\Omega>$. The simplest nontrivial diagram we are interested
in is
\begin{eqnarray} \int
\frac{dx_+dx'_+dx_-dx'_-}{\left(2\pi
i\right)^2}\frac{dy_+dy'_+dy_-dy'_-}{\left( 2\pi i\right)^2}\frac{
\hat a^{\dag }_{R,+} \left( x _+ \right) \hat b^{\dag
}_{R,+}\left( x '_+ \right)}{x'_+-x _+-i\delta } { \hat
a^{\dag }_{R,-} \left( x _- \right) \hat b^{\dag }_{R,-}\left( x
'_- \right) \over x'_- -x _- -i\delta }\nonumber \\ { \hat
a^{\dag }_{L,_+}\left( y _+ \right) \hat b^{\dag }_{L,_+}\left( y
'_+ \right) \over y_+-y'_+ -i\delta } { \hat a^{\dag
}_{L,_-}\left( y _- \right) \hat b^{\dag }_{L,_-}\left( y '_-
\right) \over y_- -y'_- -i\delta }\exp [{ {\cal S}_{\rm
eff}^f\left( x_+,\dots\right)]} |F>.  \label{s10} \end{eqnarray}
Further on we will see that the terms with a smaller number
of operators give a weaker correlation than Eq.~(\ref{s10}).

The action for this  electrons-holes configuration is
\begin{equation}
{\cal S}_{\rm eff}^f\left( x,_\alpha,\dots \right) = {1\over
2}\ln{\prod_{\alpha ,\alpha '\dots}\left( x_\alpha-y_\beta +i\delta
\right) \left( x'_{\alpha'}-y'_{\beta' }+i\delta \right)\over
\prod_{\alpha ,\alpha '\dots} \left( x'_\alpha-y_\beta +i\delta
\right)\left( x_{\alpha'}-y'_{\beta'} +i\delta \right)}.
\label{s11}
\end{equation}
In fact, it differs from the corresponding expression for
one-component fermions by the factor $1/2$. This factor leads
to the singularities of the integrand Eq.~(\ref{s10}) which are
cuts instead of simple poles in one component  case. This
prevents us from explicit calculation of the integral.

Nevertheless, it is possible to recognize the spontaneous
breakdown of the chiral symmetry in our system. To do this,
several steps are necessary. First, we have to analyze what new
bound complexes appeared as a result of interaction. We have to
take the arbitrary connected diagram and try to separate complexes
with a smaller number of particles out of it. To do this one has
to consider all particles in one complex as being one close to
another, while the distances between different complexes are
large. If the full wave function in this limit decays into the
product of two wave functions, one depending only on the
coordinates of first complex, the other only on coordinates of the
second one, then the complexes can be considered as new "free
particles" \cite{par}, as the probability to find one such complex
does not depend on the position of another one. In other words, we
should  present a term of expansion of evolution operator we
consider as a product of the form:
$$ \int {dx_+dy_+\over2\pi
i}\dots\hat a^{\dag }_{R,+}\left( x_+  \right)\dots K\left(
 x_+,\dots,y_+\dots;\right) a^{\dag }_{L,+}\left( y_+  \right)\dots\,.
$$
Then we should check that provided the complex ($x_+,\dots,$) is
moved off from the complex ($y_+\dots;$) over the distance of the
order of $L_{\Vert}$ the amplitude $K\left(
x_+,\dots,y_+,\dots\right)$ tends not to zero (as is usually the
case with scattering amplitudes) but to the factorized product $
k\left( x_+,\dots,\right)k_1\left( y_+,\dots\right)$. Here each
factor depends on the variables of the first or second group. This
means that two complexes are formed as a result of the
interaction. If the intercomplex distance is large enough, their
contribution to the wave function can be presented as follows
$$
\int {dx_+ \over \sqrt{2\pi i}}\dots k\left(
x_+\dots,\right)\hat a^{\dag}_{R,+}\left(x_+\dots\right)
\int {dy_+ \over \sqrt{2\pi i}} k_1\left( y_+\dots,\right)\hat
a^{\dag }_{L,+}\left( y_+\dots  \right)|F> .
$$
The remnant part of $K$ (which is $K-k\cdot k_1$) is a connected
diagram which describes the  intercomplex scattering. The theorem
of logarithm connectedness~\cite{FS} guarantees that the same
connected complexes will appear in the next orders as well with
correct combinatorial coefficients and the final answer is an
exponent of connected complexes. In particular, the first order
term
$$ \int {dx_+\over \sqrt{2\pi i}}\dots
k\left( x_+,\dots\right) \hat a^{\dag}_{R,+}\left( x_+\right)\dots |F>
$$
should appear in the expansion of evolution operator directly,
unless it is forbidden by some conservation law (e.g. chirality
conservation for the lowest temperature case in our model). In
this case one has to use the projector on the proper state as in
(Eq.\ref{s18})).

Whether a phase transition of the second kind exists or not,
depends on the symmetry of the complexes $\hat a^{\dag
}_{R,+}\left( x_+\right)\dots$ If they are less symmetric than the
initial Hamiltonian one has a symmetry breaking. As a result it is
possible to introduce a non-vanishing order parameter in the less
symmetric phase while it vanishes in the more symmetric one where
the complexes do not exist. (More precisely, taking into account a
fluctuations of the low symmetry phase in phase with non-broken
symmetry one can state that an order parameter should not enhance
with $L_{\Vert}$ in the last case. Our definition the order
parameter is given in Eq.(\ref{s21}).) According to Landau (see
Landau and Lifshitz~\cite{LY}) this is the definition of the phase
transition of the second kind. If, however, the symmetries of all
the connected complexes and of the Hamiltonian are the same one
has a long-range correlations without spontaneous symmetry
breaking.

Now we embark on analysis of the simplest diagrams of the
evolution operator. By analogy with the theory of one
component fermions one could assume that the simplest
connected diagrams originate from the term
\begin{equation}
\int {dx_+dx'_+\over2\pi i}{dy_-dy'_-\over 2\pi i}{ \hat a^{\dag }_{R,+}
\left( x_+  \right) \hat b^{\dag }_{R,+}\left( x'_+ \right)
\over x'_+  - x_+ -i\delta }
{ \hat a^{\dag }_{L,-}\left( y_-  \right) \hat b^{\dag }_{L,-}
\left( y '_- \right) \over y_-- y'_- -i\delta }  \exp
[{{\cal S}_{\rm eff}^f\left( x_+,\dots\right) }]|F>.
\label{s12}
\end{equation}
However, because of the factor $n^{-1}$ the corresponding contribution
to the action is
\begin{equation}
{\cal S}_{\rm eff}^f \left(  x_+,\dots \right) =
{1\over 2}\ln{{\left( x_+-y_- +i\delta \right)\left( x'_+ - y'_-
+i\delta \right)\over \left( x'_+ - y_- +i\delta
\right)\left( x_+-y'_- +i\delta \right)}}.
\label{s13}
\end{equation}

The bound chiral complexes are determined by the singularities
of the integrand at
$ |x'_+ -  y_-|\sim d, |x_+ - y'_-| \sim d,
|x_+-y_-| \sim L_{\Vert}$ (here $d$ is the width of the conductor).
As a result, the contribution we are interested in is of the order
$$
\int dx_+dy_ -\hat a^{\dag }_R\left( x_+  \right)
\hat b^{\dag }_L\left( x_+ \right)
\hat a^{\dag }_L\left( y_-  \right)  \hat b^{\dag }_R
\left(  y_- \right){d\over |x_+-y_-|}|F>.
$$
This quantity tends to 0 at $|x_+-y_-| \to \infty $, but more
slowly than for free particle~\cite{rev1} case.

Now we will show that the most correlated state can be obtained
from Eq.(~\ref{s10}). It can be split into two four-particle
complexes, each with zero spin, having the chiral charges $\pm 4 $
($\hat a^{\dag }_{R,+}\hat a^{\dag }_{R,-}\hat b^{\dag }_{L,_+}
\hat b^{\dag }_{L,_-}$ and $\hat a^{\dag }_{L,_+} \hat a^{\dag
}_{L,_-} \hat b^{\dag }_{R,+}\hat b^{\dag }_{R,-}$). The amplitude
$K$ in Eq.(~\ref{s10}) factorizes and does not tend to 0 at
$L_{\Vert} \to \infty .$ Indeed, the $c$-factor in the integrand
is
 \begin{eqnarray}
K\left( x_+,\dots  \right)={ 1 \over x'_+-x _+-i\delta }
{1 \over x'_- -x _- -i\delta } { 1 \over y_+-y'_+ -i\delta }
{ 1 \over y_- -y'_- -i\delta }\nonumber\\
{\sqrt {\prod_{\alpha ,\alpha '\dots}\left( x_\alpha-y_\beta +i\delta
\right)
\left( x'_{\alpha'}-y'_{\beta' }+i\delta \right)}
\over \sqrt {\prod_{\alpha ,\alpha '\dots}
\left( x'_\alpha-y_\beta +i\delta \right)
\left( x_{\alpha'}-y'_{\beta'} +i\delta \right)}}.
\label{s10a}
\end{eqnarray}

Now, let us consider the regions of integration
$x_+ \sim x_- \sim y'_- \sim y'_+$
and $x'_+\sim x'_- \sim y_+ \sim y_-$
assuming that the distances between these groups of variables
are of the order of $L_{\Vert}$. Then the amplitude $K$ tends to
\begin{eqnarray}
V_{+4}\left( x_+,\dots  \right)V_{-4}\left( x'_+,\dots  \right)=\nonumber\\
=1/\sqrt {\left( x_+ - y'_++i\delta \right)\left( x_+ - y'_-+i\delta \right)
\left( x_- - y'_++i\delta \right)\left( x_- - y'_-+i\delta \right)}\nonumber\\
1/\sqrt {\left( x'_+ - y_++i\delta \right)\left( x'_+ - y_-+i\delta \right)
\left( x'_- - y_++i\delta \right)\left( x'_- - y_-+i\delta \right)},
\label{s14}
\end{eqnarray}
This means that each amplitude $V$  depends on the variables
belonging either to the first or to the second group. This
property of the amplitude permits one to single out of the full
equation for the evolution operator the connected complexes and
the amplitude of intercomplex scattering that tend to 0 for
large intercomplex distances.

This term of expansion, besides the chiral complexes, has also
neutral ones with zero chirality, $\hat a^{\dag }_{R,+}\hat b^{\dag
}_{R,+} \hat a^{\dag }_{L,_-}\hat b^{\dag }_{L,_-}$  and $\hat
a^{\dag }_{R,-}  \hat b^{\dag }_{R,-} \hat a^{\dag }_{L,_+}\hat
b^{\dag }_{L,_+}$. They do not violate the symmetry of the
Hamiltonian. However, they should be singled out, so that the
remaining scattering amplitude tended to zero in the whole region
of the variable variation. This permits one to interpret it as the
intercomplex scattering amplitude. The zero chirality complexes
are not important for existence of the phase transition. However,
one should take them into account for calculation of the matrix
elements as the theory has no small parameter to
neglect them. To
check that they exist we consider in~(\ref{s10}) the region $x_+
\sim x'_+\sim y_- \sim y'_-$ Ё $x_-\sim x'_- \sim y_+ \sim y'_+$.
In this region the amplitude $K$ tends to $V_0\left( x_+ ,\dots
\right) V_0\left( x_- ,\dots  \right)$ where
\begin{eqnarray}
V_0\left( x_\alpha ,\dots  \right)={1\over
\left(x'_\alpha -x _\alpha-i\delta  \right)\left(y_{-\alpha}-y'_{-\alpha}
-i\delta  \right)}
{\sqrt {\left( x_\alpha-y_{-\alpha} +i\delta \right)
\left( x'_{\alpha}-y'_{-\alpha }+i\delta \right)}\over
\sqrt {\left( x_\alpha-y'_{-\alpha } +i\delta \right)
\left( x'_{\alpha}-y'_{-\alpha } +i\delta \right)}}.
\label{s15}
\end{eqnarray}
This means that this quantity can be presented as a product of the
amplitudes, each of them remaining finite provided the distance
between them tends to infinity.

Now it is convenient to introduce the amplitude of intercomplex
scattering $V_{coll}$. By derivation, it tends to zero, the
intercomplex distance tending to $\infty$:
\begin{equation}
V_{coll}\left( x_+ ,\dots  \right) = K\left( x_+ ,\dots \right) -
V_{+4}\left( x_+,\dots  \right)V_{-4}\left( x'_+,\dots \right) -
V_0\left( x_+ ,\dots  \right) V_0\left( x_- ,\dots  \right) .
\label{s16}
\end{equation}
The contribution to the  ground state wave function we are interested
in can be presented through these amplitudes as
\begin{eqnarray}
\int {dx_+dx'_+dx_-dx'_-\over\left(2\pi i\right)^2}
{dy_+dy'_+dy_-dy'_-\over \left( 2\pi i\right)^2}
\hat a^{\dag }_{R,+}
\left( x _+ \right) \hat b^{\dag }_{R,+}
\left( x '_+ \right)  \hat a^{\dag }_{R,-}
\left( x _- \right) \nonumber \\
\hat b^{\dag }_{R,-}\left( x '_- \right)
\hat a^{\dag }_{L,_+}\left( y _+ \right)
\hat b^{\dag }_{L,_+}\left( y '_+ \right)
 \hat a^{\dag }_{L,_-}\left( y _- \right)
\hat b^{\dag }_{L,_-}\left( y '_- \right) \nonumber \\
\left(  V_{+4}\left( x_+,\dots  \right)V_{-4}\left( x'_+,\dots
\right) + V_0\left( x_+ ,\dots  \right) V_0\left( x_- ,\dots
\right) +V_{coll}\left( x_+ ,\dots  \right) \right) |F>.
\label{s17}
\end{eqnarray}
First term here describes the non-interacting complexes with
non-zero chirality, second describes the 4-particle neutral
complexes and third (connected part) is related to their
interaction. We are mainly interested in first term as it is
connected with breakdown of chiral symmetry.

The chiral complex which we obtained is already a connected one
and cannot be separated to simpler ones. This means that its
wave function is a decreasing function of interparticle distances.
It is shown in the Appendix that probability to find particles
of the complex far from each other is negligibly small.

Besides, one should take into consideration that in the
temperature interval $\Theta\ll \Theta_{\rm chiral}$
where one need not consider the last term in Eq.~(\ref{s7})
there is one to one correspondence between the complexes
with chirality $Q=4$ and $Q=-4$, so that the total
chirality of the state equals zero. The theorem of logarithm
connectedness states that the ground state wave function can be
presented as
\begin{eqnarray}
|\Omega>
=P_{C=0} \exp \sum_\alpha {\rm Tr} \left[  {1\over\left( 2\pi
i\right)^2} V_{+4}\left( x_\alpha,  x_{-\alpha}, y'_{-\alpha} ,
y'_{\alpha}\right)\left( \hat a^{\dag }_{R,\alpha}\left(
x_\alpha\right)\hat a^{\dag }_{R,-\alpha}\left(
x_{-\alpha}\right)\right. \right.  \nonumber\\
\times\left. \left.\hat
b^{\dag }_{L,-\alpha}\left(  y'_{-\alpha}\right)\hat b^{\dag
}_{L,\alpha} \left(  y'_{\alpha}\right) + \hat a^{\dag
}_{L,\alpha}\left( y'_\alpha\right)\hat a^{\dag
}_{L,-\alpha}\left(  y'_{-\alpha}\right) \hat b^{\dag
}_{R,-\alpha}\left(  x_{-\alpha}\right)\hat b^{\dag }_{R,_\alpha}
\left(  x_{\alpha}\right)\right) \right.\nonumber\\
\left.
+{1\over\left( 2\pi i\right)^2} V_{+0}\left( x_\alpha,
x_{-\alpha}, y'_{-\alpha} , y'_{\alpha}\right)\left( \hat a^{\dag
}_{R,\alpha}\left( x_\alpha\right)\hat b^{\dag }_{R,\alpha}\left(
x'_{\alpha}\right) \right. \right.  \nonumber\\
\times\left. \left.\hat
a^{\dag }_{L,-\alpha}\left(  y_{-\alpha}\right)\hat b^{\dag
}_{L,-\alpha} \left(  y'_{-\alpha}\right) + \hat a^{\dag }_{R,
-\alpha}\left( x_-\alpha\right)\hat b^{\dag }_{R,-\alpha}\left(
x'_{-\alpha}\right) \hat a^{\dag }_{L,\alpha}\left(
y_{\alpha}\right)\hat b^{\dag }_{L,\alpha} \left(
y'_{\alpha}\right) \right) \right.\nonumber\\
 \left. +{1\over
\left( 2\pi i\right)^4} V_{coll}\left( x_\alpha,  x_{-\alpha},
...\right)\hat a^{\dag }_{R,\alpha}\left( x_\alpha\right) \hat
a^{\dag }_{R,-\alpha}\left(  x_{-\alpha}\right) \hat b^{\dag
}_{L,-\alpha}\left(  y'_{-\alpha}\right)\hat b^{\dag }_{L,\alpha}
\left(  y'_{\alpha}\right) \right.\nonumber\\
\times\left.\hat a^{\dag
}_{L,\alpha}\left( y_\alpha\right)\hat a^{\dag
}_{L,-\alpha}\left(  y_{-\alpha}\right) \hat b^{\dag
}_{R,-\alpha}\left(  x'_{-\alpha}\right)\hat b^{\dag
}_{R,_\alpha} \left(  x'_{\alpha}\right) +\dots  \right].
 \label{s18}
\end{eqnarray}
Here $P_{C=0}$ is the projector on the state with zero chirality.
The symbol $\rm Tr$ includes the integrations in over particle
coordinates. The terms omitted in Eq.~(\ref{s18}) describe
scattering of three and more complexes while all the elementary
complexes are present here. Mark that the complexes with non-zero
chirality have appeared in the theory. Nevertheless, the wave
function of the ground state as a whole described the state with
$Q=0$ in this temperature region, i. e. the symmetry of the ground
state is the same as that of the Hamiltonian. The states with
nonzero chirality have a bigger energy (of the order of $2\pi
v_F/L_{\Vert}$). Therefore the spontaneous symmetry breaking
 may take place only in the region of higher temperatures $\Theta\gg
\Theta_{\rm chiral}$  where such an energy difference is not
essential. In this temperature region one should consider in the
equation ~(\ref{g7}) for the action ${\cal S}_0 $ also the term
where the time arguments of the Green functions differ by $T$.
(Practically it is more convenient to introduce the temperature by
replacement $T \to -i/\Theta $ in the final equations.) Then one
has the following non-trivial term in the operator of evolution
\begin{eqnarray} {\rm Tr}
{1\over\left( 2\pi i\right)^4} {\hat a^{\dag }_{R,+} \left( x_+
\right) \hat a_R\left( \tilde x_+  \right) \over \tilde x _+- x_+
+v_fT-i\delta } {\hat a^{\dag }_{R,-} \left( x_-  \right) \hat
a_R\left( \tilde x_-  \right) \over \tilde x _-- x_-
+v_fT-i\delta } \nonumber\\ { \hat b^{\dag }_{L,+}\left( y'_+
\right) \hat b_L\left( \tilde y '_+ \right) \over \tilde y'_+ -
 y'_+ -v_fT+i\delta } { \hat b^{\dag }_{L,_-}\left( y'_-  \right)
 \hat b_L\left( \tilde y '_- \right) \over \tilde y'_- - y'_-
-v_fT+i\delta } \exp [{{\cal S}_{\rm eff}^f]\left( x_+,\dots\right) }
|F><F|.  \label{s19} \end{eqnarray}
The action  ${\cal S}_{\rm eff}^f$
for this configuration is
\begin{eqnarray} {1\over
2}\ln{{\prod_{\alpha ,\alpha '...}\left( \tilde y'_\alpha -
y'_\beta -v_fT+i\delta \right) \left( x_{\alpha'}- \tilde x
_{\beta'} -v_fT+i\delta \right) \over \prod_{\alpha ,\alpha '...}
\left( x_\alpha -y'_\beta+i\delta \right) \left( \tilde
y'_{\alpha'} -\tilde x_{\beta'} +i\delta \right)}}.
\label{s19a}
\end{eqnarray}

It is readily seen out of the operator structure of this term that
the amplitude $V_4$  appears here automatically (without
extracting the amplitudes of the neutral complexes and channels of
scattering). It is a consequence of the theorem of logarithm
connectedness. It guarantees coincidence of the amplitude in this
term with $V_4$. Indeed, in the region where the same variables
with spin up and spin down are quite near to each other (for
instance, $x_\alpha \sim x_{-\alpha}$) while the coordinates in
the creation and annihilation operators are apart at the distance
of the order of $L_{\Vert}$ ($x_\alpha \sim  \tilde x_{\alpha}
\sim L \gg 2\pi v_f/\Theta$) all the $c$-factor in the integrand
of~(\ref{s19}) turns into the factorized expression
$$
V_{+4}\left( x_+,\dots
\right) V^*_{-4}\left( \tilde x_+,\dots  \right) .
$$

This proves that the chiral four particle complexes were
singled properly out of a more complicated expression~%
(\ref{s16}). Such terms in the evolution operator result
in any chirality of the ground state. However, any state
with a fixed chirality should be unstable relative to
the backscattering, however weak, violating the chirality.
Therefore one should consider a superposition of all such
states. As a result, one has, in the same way as in the
theory of superconductivity, to introduce a condensate
with a fixed phase instead of a state with a fixed
chirality:

 \begin{eqnarray} |\Omega>_{\theta} = \exp \sum_\alpha {\rm Tr}
 \left[
{1\over\left( 2\pi i\right)^2} V_{+4}\left( x_\alpha,
x_{-\alpha}, y'_{-\alpha} , y'_{\alpha}\right)\left( \exp{\left(
i\theta\right)} \hat a^{\dag }_{R,\alpha}\left(
x_\alpha\right)\hat a^{\dag }_{R,-\alpha}\left(
x_{-\alpha}\right)\right. \right.  \nonumber\\
\times\left. \left.\hat
b^{\dag }_{L,-\alpha}\left(  y'_{-\alpha}\right)\hat b^{\dag
}_{L,\alpha} \left(  y'_{\alpha}\right) +
\exp{\left(-i\theta\right)} \hat a^{\dag }_{L,\alpha}\left(
y'_\alpha\right)\hat a^{\dag }_{L,-\alpha}\left(
y'_{-\alpha}\right) \hat b^{\dag }_{R,-\alpha}\left(
x_{-\alpha}\right)\hat b^{\dag }_{R,_\alpha} \left(
x_{\alpha}\right)\right) \right.\nonumber\\
\left. +{1\over\left(
2\pi i\right)^2} V_{+0}\left( x_\alpha,  x_{-\alpha},
y'_{-\alpha} , y'_{\alpha}\right)\left( \hat a^{\dag
}_{R,\alpha}\left( x_\alpha\right)\hat b^{\dag }_{R,\alpha}\left(
x'_{\alpha}\right) \right. \right.  \nonumber\\
\times\left. \left.\hat
a^{\dag }_{L,-\alpha}\left(  y_{-\alpha}\right)\hat b^{\dag
}_{L,-\alpha} \left(  y'_{-\alpha}\right) + \hat a^{\dag }_{R,
-\alpha}\left( x_-\alpha\right)\hat b^{\dag }_{R,-\alpha}\left(
x'_{-\alpha}\right) \hat a^{\dag }_{L,\alpha}\left(
y_{\alpha}\right)\hat b^{\dag }_{L,\alpha} \left(
y'_{\alpha}\right) \right) \right.\nonumber\\
\left. +{1\over \left( 2\pi i\right)^4} V_{coll}\left( x_\alpha,
x_{-\alpha}, ...\right)\hat a^{\dag }_{R,\alpha}\left(
x_\alpha\right) \hat a^{\dag }_{R,-\alpha}\left(
x_{-\alpha}\right) \hat b^{\dag }_{L,-\alpha}\left(
y'_{-\alpha}\right)\hat b^{\dag }_{L,\alpha}
\left(  y'_{\alpha}\right) \right.\nonumber\\
\times\left.\hat a^{\dag
}_{L,\alpha}\left( y_\alpha\right)\hat a^{\dag
}_{L,-\alpha}\left(  y_{-\alpha}\right) \hat b^{\dag
}_{R,-\alpha}\left(  x'_{-\alpha}\right)\hat b^{\dag
}_{R,_\alpha} \left(  x'_{\alpha}\right) +....  \right],
 \label{s20}
 \end{eqnarray}

Eq.~(\ref{s20}) demonstrates that for a strong electron-electron
interaction a spontaneous symmetry breaking takes place. Two first
terms in the equation for the ground state wave function are not
invariant under the chiral transformation
\begin{equation}
\Psi_{R,L}\left( x  \right) \to \exp{\left( \pm
i\Lambda\right) }\Psi_{R,L},
\label{s20a}
\end{equation}
while the Hamiltonian still retains the invariance. (Here $\Lambda$ is a
constant.)

Such a form of the bound state permits to introduce the order
parameter. It equals to zero (or, to be more exact, not grows with
$L_{\Vert}$) in the phase of high symmetry ($\Theta \gg \Theta_c =
\omega_{p_{\rm min}}$) and is proportional to $L_{\Vert}$ in the
phase of low symmetry for low temperatures. (One can prove the
first statement using ordinary symmetry considerations or by
direct analytical calculation.) For our case the following
quantity can be considered as the order parameter
\begin{equation}
\int dx \left.\right._{\theta}<\Omega | \hat
a^{\dag }_{R,\alpha}\left( x \right)\hat a^{\dag
}_{R,-\alpha}\left(  x \right) \hat b^{\dag }_{L,\alpha}\left(  x
\right) \hat b^{\dag }_{L,-\alpha}\left(  x \right)
|\Omega>_{\theta} \sim L_{\Vert} .
\label{s21}
\end{equation}

One can see from Eq.~(\ref{s21}) as to why the phase transition of
the second kind demands the chirality degeneracy of the ground
state. For the order parameter to be nonvanishing it is necessary
to be able to add to the ground state an extra four particle
complex. This demands  condition $\Theta\gg\Theta_{\rm chiral}$
because  zero chirality state has the lowest energy and the energy
difference between the state  and other ones about $\Theta_{\rm
chiral}$. There is also the upper bound of existence of the low
symmetry phase
\begin{equation}
\Theta\ll \omega_{p_{\rm min}}.
\label{s22}
\end{equation}
This limitation is due to the fact that in one dimensional
systems the long range order is suppressed by the thermal
excitations. In the temperature region given by
Eq.~(\ref{s22}) one can neglect it; this condition does not depend
on the number of the fermion components and is discussed
in detail in~\cite{AP}. So, the temperature region
\begin{equation}
\Theta_{\rm chiral} < \Theta < \Theta_c
\label{s8t}
\end{equation}
is the region where the broken phase exists.

Thus in the system of interacting multicomponent fermions the most
correlated state consists of $2n$ operators and has the chirality
$\pm 2n$.  It is this state that in the limit of infinitely strong
interaction results in the phase transition of the second kind.
At the same time the spin complexes of a smaller number of
operators, violating
the symmetry of the ground state, can exist only as a Kosterlitz
--- Thouless phase. For the limit of infinitely
strong interaction their density tends to zero
as $1/\sqrt{L_{\Vert}}$.
By contrast, the spinless
phase has  for this case a finite density. This
rule is quite general.

\section{carbon nanotubes}\label{CN}
The conducting carbon nanotubes give one more example of
multicomponent electrons. In order to generalize the theory developed
above one should have one dimensional conducting tubes and such e-e
interaction that could be rewritten in the density-density
form --- see Eq.(\ref{g1}). It means that one should be able to
neglect a backward and inter-component (see below) scattering.
Following Ref.~\onlinecite{AA} one can visualize a nanotube as a
cylinder constructed of a monoatomic layer of graphite. The latter has
a lattice of adjoining regular hexagons, so that the angle between the
neighboring basis vectors, $n_{a}$ and $n_{b}$ is $2\pi/3$. Choosing
the coordinates $\xi_1$ and $\xi_2$ in such a way that axis 0$\xi_1$ is
parallel to ${\bf a}$ while axis 0$\xi_2$ is perpendicular to ${\bf a}$
one can present these vectors as
\begin{equation}
{\bf a}=a(1,0),\quad
{\bf b}=a(-1/2,\sqrt{3}/2).
\label{2n}
\end{equation}
$a$ is the lattice constant that is equal to $d\sqrt{3}$,
$d=$1.44 \AA \, being the interatomic distance~\cite{JS}.

The circumferential vector $\bf L$ can be written as
\begin{equation}
{\bf L}=n_a{\bf a}+n_b{\bf b}.
\label{1n}
\end{equation}
Here $n_a$ and $n_b$ are integers.

The electron effective Hamiltonian for a graphite sheet is
\begin{equation}
H=\left({0\quad h^*}\atop{h\quad0}\right).
\label{3n}
\end{equation}
It can be expanded in the vicinity of the points
\begin{equation}
{\bf P}=(4\pi/3a)(-1,0),\quad {\bf P'}=(4\pi/3a)(1,0)
\label{4n}
\end{equation}
up to the first power in the small deviations $\bf p$ and $\bf p'$ from
the values given by first and second Eq.~(\ref{4n}) respectively
\begin{equation}
h({\bf P},{\bf p})=\gamma e^{-i\theta}\left(p_{\perp}-ip_z\right),\quad
h({\bf P'},{\bf p'})=\gamma e^{i\theta}\left(-p'_{\perp}-ip'_z\right).
\label{5n}
\end{equation}
Here $\gamma=(\sqrt{3}/2)\gamma_0a;\,\gamma_0\approx\,
$3eV~(see~\onlinecite{AA,JS}) is the transfer integral
between the neighboring $\pi$ orbitals while $\theta$
is the angle between vectors $\bf L$ and $\bf a$. The
subscripts $z$ and $\perp$ refer to the components of
$\bf p$ relative to the direction of $\bf L$, namely
$p_z\perp{\bf L}$ and $p_{\perp}\Vert{\bf L}$, so that
$p_z (p_{\perp})$ is parallel (perpendicular) to the axis
of the tube.

The spectrum near the point $\bf P$ is given by
\begin{equation}
E({\bf P},{\bf p})=\pm\gamma\sqrt{p_z^2+p_{\perp}^2}
\label{6n}
\end{equation}
where the upper (lower) sign corresponds to the conduction
(valence) band in this equation. The spectrum near the $\bf P'$
point is obtained by the replacement ${\bf p}\to{\bf p'}$.

The electron wave function $\Psi({\bf r})$ should satisfy the
cyclic boundary condition
\begin{equation}
\Psi({\bf r})=\Psi({\bf r}+{\bf L}),
\label{7n}
\end{equation}
so that the discrete values of $p_{\perp}$ and $p'_{\perp}$
are given by (see Ref.~\onlinecite{AA})
\begin{equation}
p_{\perp}={2\pi\over \vert {\bf L}\vert}\left(n-{\nu\over3}\right);\quad
p'_{\perp}={2\pi\over \vert {\bf L}\vert}\left(n+{\nu\over3}\right) .
\label{8n}
\end{equation}
Here $n=0,\pm1,\pm2,\dots$ while $\vert {\bf L}\vert=a\sqrt{n_a^2+n_b^2
-n_an_b}$; $\nu=0\; \mbox{or}\, \pm1$ and is determined by the
presentation of the sum $n_a+n_b$ as $3N+\nu$ ($N$
being an integer). The nanotubes are conductive (metallic) for
the combination
\begin{equation}
n=\nu=0
\label{9n}
\end{equation}
and we will consider this case for our further analysis. In other
words, in such tubes there are two conic bands,
i. e. the points $\alpha_P{\bf P}$ with
$\alpha_P =\pm1$. The big phase corresponding to the momentum $\alpha_P{\bf P}$
should be extracted in the same way as it has been done in
Eq.~\ref{g2}. Besides, we assume that due to the presence of gate
electrodes the Fermi level is well above (or below) the points
$\alpha_P{\bf P}$ (cf. with Ref.~\cite{NT}). As a result, we will
have a theory with four-component fermions. There are two extra
branches corresponding to two values of $\alpha_P$. In each of them
there are analogues of $R$- and $L$-particles. To introduce them one
should, in full analogy with Eq.~(\ref{g2}), separate the
corresponding phase factors with large phases. In order to
establish correspondence between the present model and the Luttinger
one we should be able to neglect both the transitions between different
branches (different values of $\alpha_P$) and within the same branch
between $R$- and $L$-particles.
As indicated in Ref.~\onlinecite{KBF}, the nanotubes have comparatively
large radii that encompass with ($N\gg1$) atoms. Therefore the only e-e
scattering that is important in this limit is the forward one that
involves a small quasimomentum transfer. The matrix element describing
the
backscattering within a band as well as the ${\bf P}\leftrightarrow{\bf
P'}$ scattering acquire an extra small factor of the order of $1/N$.
This is why one can neglect these types of scattering.  This means that
we can use the results obtained in Section~\ref{G}.  Repeating the
arguments of this section for $n=4$ one can come to the conclusion that
a condensate is formed in the ground state. It consists of the eights
of the form
$$ \hat a^{\dag }_{R,\alpha ,\alpha_P }\left( x \right)\hat
a^{\dag }_{R,-\alpha,\alpha_P}\left(  x \right) \hat b^{\dag
}_{L,\alpha ,\alpha_P}\left(  x \right) \hat b^{\dag
}_{L,-\alpha,\alpha_P}\left(  x \right) \hat a^{\dag }_{R,\alpha
,-\alpha_P }\left( x \right)\hat a^{\dag
}_{R,-\alpha,-\alpha_P}\left(  x \right) \hat b^{\dag }_{L,\alpha
,-\alpha_P}\left(  x \right) \hat b^{\dag
}_{L,-\alpha,-\alpha_P}\left(  x \right),
$$
Their chirality is $\pm 8$.

\acknowledgments

V. V. A. and V. L. G. are grateful for partial support the work by
the Russian National Foundation for Basic Research, grant No
06-02-16384.

\appendix
\section{Characteristic dimensions of correlated complexes}
The simplest way to give the proof of existence of a bound chiral
complex and to determine its characteristic dimensions is to
consider the state with a single chiral complex:
\begin{equation}
|\Phi_c>={\rm Tr}
\left[\frac{\hat a^{\dag }_{R,+}( x_+)\hat
a^{\dag }_{R,-}(  x_{-}) \hat b^{\dag }_{L,-}(
y'_{-})\hat b^{\dag }_{L,+} (
y'_{+})}{\sqrt{ \prod_{\alpha ,\alpha' = \pm
}( x_\alpha-y'_{\alpha'} + i\delta )(
x_\alpha-y'_{\alpha'} + i\delta )}}\right] |F> .
\label{a1}
\end{equation}
For instance, one can calculate the probability to find
an electron with spin up at the distance $|z_+ - z_-|> d$ from the electron
with spin down as
$$
 <\Phi_c| \hat a^{\dag }_{R,+}\left( z_+\right) a_{R,+}\left( z_+\right)
\hat a^{\dag }_{R,-}\left(  z_{-}\right)\hat a_{R,-}\left(  z_{-}\right)|\Phi_c> .
$$

Moving all the creation operators to the right and all the
annihilation operators to the left one gets  $A^2/|z_+-z_-|^2$.
Here $A$ is a constant equal to
$$
 \int dx \left( \left( 1+x\right)^2+ \delta^2\right)^{-1/2}
\left( \left( 1-x\right)^2+ \delta^2\right)^{-1/2}.
$$
This means that the most probable is the particle configuration
where $|z_+-z_-|\sim d$ (to get the physical parameter one
should replace, as usual, $\delta$ by $d$).
That is  the right and left electrons with
the opposite spins are in fact always near one another forming a
chiral complex. In this sense the chiral fours are point-like
entities as the $R\overline{L}$-pairs for one component fermions.
In the same way one can give estimates of the distances between
all the particles belonging to a four-particle complex.

The same calculation for a neutral four-particle complex are a
little more cumbersome. The state with a single neutral pair
is described by the wave function
\begin{eqnarray}
|\Phi_0>={\rm Tr}\left[ \hat a^{\dag }_{R,-}\left( x_-\right)\hat
b^{\dag }_{R,-}\left(  x'_{-}\right)
 \hat a^{\dag }_{L,+}\left(  y_{+}\right)
\hat b^{\dag }_{L,+}
\left(  y'_{+}\right)
\right.\nonumber\\
\left. \left( x'_- -x_- -i\delta\right)^{-1} \left( y_{+}-y'_{+}-
i\delta\right)^{-1}
{ \sqrt{ \left(   x_- -y_{+} + i\delta \right)
\left(   x'_- -y'_{+} + i\delta \right)  }\over
\sqrt{ \left(   x_- -y'_{+} + i\delta \right)
\left(   x'_- -y_{+} + i\delta \right)  } } \right]|F>.
\label{a2}
\end{eqnarray}
In order to find the characteristic size of a neutral complex
consider the matrix element
$$
<\Phi_0| \hat a^{\dag }_{R,-}\left( z_-\right) a_{R,-}\left( z_-\right)
\hat a^{\dag }_{L,+}\left(  z_{+}\right)\hat a_{L+}\left(  z_{+}\right)|\Phi_0>
$$
It is equal to
\begin{eqnarray}
{\rm Tr'}\left[ \left( \tilde x'_- -\overline {\tilde x' }_- -i\delta\right)^{-1}
\left( \overline {\tilde y'}_{+}-\tilde y'_{+}-i\delta\right)^{-1}
\left( \overline {\tilde x'}
_- +i\delta\right)^{-1}
\left( -\overline {\tilde y'}_{+} +i\delta\right)^{-1}\right.\nonumber\\
\left.{ \sqrt{ \left(  z_- - z_{+} - i\delta \right)
\left(  z_- -z_++\overline{ \tilde x'}_- -\overline{\tilde y'}_{+} - i\delta \right)  }\over\sqrt{ \left(   z_- -z_+ -
\overline{\tilde y'}_+ - i\delta \right)
\left(   z_ - - z_+ + \overline{\tilde x'}_-  - i\delta \right) } } \right.\nonumber\\
 \left.\left( \tilde x'_- - i\delta\right)^{-1}
\left( -\tilde y'_{+}-i\delta\right)^{-1}{ \sqrt{ \left(   z_- -z_{+} + i\delta \right)
\left(   z_- - z_+ +\tilde x'_- -\tilde y'_{+} + i\delta \right)
}\over\sqrt{ \left(  z_- - z_+ - \tilde y'_{+} + i\delta \right)
\left(  z_- - z_+ +\tilde x'_-  + i\delta \right)  } }\right]
\label{a3} \end{eqnarray}
($\rm Tr'$ implies that one should
integrate over all the variables besides $z_\alpha$). The exact
expression for this matrix element for arbitrary values of $\Delta
Z=z_+-z_-$  is rather cumbersome and non-informative.
It is sufficient to prove that the most probable is the
particle configuration where $ \Delta Z\sim d $.

To do this let us note that for $ \Delta Z\sim \delta $
all the integrals in Eq.~({a3}) converge and are dominated
by the regions of the order $\delta $. Let us show that
for $ \Delta Z\gg \delta $ the matrix element (\ref{a3})
has an additional small factor $\delta /\Delta Z$.
Consider, for instance, the integration
over $\tilde x'_-$.  Only the factor
\begin{equation} { \sqrt{ \left(   z_- - z_+ +\tilde
x'_- -\tilde y'_{+} + i\delta \right)  } \over\sqrt{ \left(  z_-
- z_+ +\tilde x'_-  + i\delta \right)  } }.  \label{a4}
\end{equation}
in the integrand has a singularity in the lower semiplane.
The rest integrand does not depend of $ \Delta Z$ and has
singularities only in the upper semiplane. In the main
approximation in $\delta/\Delta Z$ (\ref{a4}) tends to 1.
($ \Delta Z\gg \delta ,$ while the regions $\tilde x'_-,\tilde
 y'_{+}$, giving the main contribution into the integral
$\sim \delta $). Hence in this approximation all the integral
(\ref{a3}) vanishes. It is nonzero only in the next approximation
due to the factors of the sort $\sqrt{ \left( \tilde x'_-  +
i\delta \right)/  \left( z_- - z_+ \right) }\ll 1$. Thus the
probability  to find the electrons we are interested in at
a large distance is small. Most probable is a four where these
particles are at the distances $\sim d$.



\begin{thebibliography}{99}
\bibitem{V}J. Voit, Rep. Prog. Phys. {\bf57}, 977 (1994).

\bibitem{E}V. J. Emery, in {\em "Highly Conducting One-Dimensional Solids}",
ed. J. T. Devreese et al. (Plenum, New York, 1979), p.327.

\bibitem{Shul}
H. J. Schulz, Phys. Rev.Let. {\bf71}, 1864 (1993).

\bibitem{AP}
V.V. Afonin and V.Yu. Petrov, cond-mat/0407351

\bibitem{OD}
A. A. Odintsov and  Hideo Yoshika, Phys. Rev.B {\bf 59}, R10457
(1999).

\bibitem{NT}
R. Egger and  A. O. Gogolin, Phys. Rev. Let. {\bf 79}, 5082  (1997).

\bibitem{K} C. Kane, L. Balents, and M. P. A. Fisher, Phys. Rev.
Let. {\bf79}, 5086 (1997).
\bibitem{thooft}
G't Hooft, Phys. Rev. D {\bf14} p.3432 (1976)
\bibitem{smilga}
A.V. Smilga, Phys. Rev. D {\bf46}, 5598 (1992)
\bibitem{M}G.D.Machan, {\em"Many Particle Physics"} Plenum Press, New
York, (1993).

\bibitem{Hub}J. Hubbard,  Phys. Rev. Let. {\bf3},
77 (1959).

\bibitem{TS}
Alexei M.Tsvelik, {\em"Quantum Field Theory in Condenced Matter Physics",}
Cambridge, University Press (1998).

\bibitem{FS} See, for example, Slavnov A. A, Faddeev L. D.
{\em"Introduction to the quantum field theory".} Moskow, "Nauka" (1978).
This theorem is, in fact, a purely combinatorial statement. In the
field theory we apply it mostly to Green functions. In statistical
physics it is known as the first Mayer's theorem (Uhlenbek G.E., Ford
G.W. and Montroll E.W. {\em"Lectures in statistical mechanics,"}
American Mathematical Society, Providence  (1963)).

\bibitem{par} Note that the analogy with a bound state is quite limited.
It would be more correct to write about a correlation in the momentum
space.

\bibitem{LY}
L. D. Landau, E. M. Lifshitz, {\em "Statistical Physics"}, Pergamon,
1986.

\bibitem{rev1}
This means  that even  in the  strong interaction  limit the  spin
phase  can  exist  as  a  Kosterlitz  ---  Thouless phase with the
Thouless     constant      of      the     order      of      1/2.
We  assign  the  chirality  $+1$  to  a  right electron and a left
hole       and       $-1$       to       their       counterparts.

\bibitem{AA} H. Ajiki and T. Ando, J. Phys. Soc. Jpn. {\bf 62}, 1255
(1993).

\bibitem{JS}A. Jorio, R. Saito, C. M. Lieber, M. Hunter, T.
McClure, G. Dresselhaus, and M. S. Dresselhaus, \prl {\bf86},
1118 (2001).

\bibitem{KBF}
C.Kane, L.Balents, and P.A.Fisher
Phys. Rev. Let. {\bf79}, 5086 (1997).
\end{thebibliography}
\end{document}